\documentclass[12pt,aps,showpacs,prd,nofootinbib]{revtex4-1}

\usepackage{amssymb,amsthm,amsmath}
\newcommand{\R}{{\mathbb R}}
\newcommand{\C}{{\mathbb C}}

\newcommand{\im}{{\rm i}}

\newcommand\be{\begin{eqnarray}}
\newcommand\ee{\end{eqnarray}}

\begin{document}

\title{A Unified Theory of \\ Non-Linear 
Electrodynamics and Gravity}
\author{Alexander Torres-Gomez, Kirill Krasnov and Carlos Scarinci}
\affiliation{School of Mathematical Sciences, University of Nottingham, Nottingham, NG7 2RD, UK}
\date{November 2010}

\begin{abstract} We describe a class of unified theories of electromagnetism and gravity. The Lagrangian is of
the BF type, with a potential for the B-field, the gauge group is ${\rm U}(2)$ (complexified). Given a choice of the potential function the theory is a deformation of (complex) general relativity and electromagnetism, and describes just two
propagating polarisations of the graviton and two of the photon. When gravity is switched off the theory becomes
the usual non-linear electrodynamics with a general structure function. The Einstein-Maxwell theory can be recovered by 
sending some of the parameters of the defining potential to zero, but for any generic choice of the potential the theory is indistinguishable from Einstein-Maxwell at low energies. A real theory is obtained by imposing suitable reality 
conditions. We also study the spherically-symmetric solution and show how the usual Reissner-Nordstrom solution is recovered. 
\end{abstract}

\pacs{}

\maketitle

\section{Introduction}

In the seventies Plebanski proposed \cite{Plebanski:1977zz} a reformulation
of general relativity where the basic dynamical object was taken to be a collection of 
two-forms satisfying certain algebraic constraints. Essentially the same formulation was rediscovered a decade later
via a completely different path (canonical transformation on the phase space of GR) by
Ashtekar \cite{Ashtekar:1987gu}. A relation between the new Hamiltonian formulation
\cite{Ashtekar:1987gu} and Plebanski's theory was
elucidated in \cite{Jacobson:1988yy} and works that followed, in particular
\cite{Capovilla:1991qb}. These developments have made it clear that general relativity
can be formulated as a theory whose phase space is that of pairs
(connection, canonically conjugate electric field), i.e. as a
gauge theory. This immediately suggested that it might be possible to unify gravity
with other interactions by considering a version of Plebanski-Ashtekar theory written for
a gauge group larger than (complexified) ${\rm SU}(2)$ that gives  GR.
This unification program has been pursued in \cite{Peldan:1992iw}, \cite{Chakraborty:1994vx},
\cite{Chakraborty:1994yr} at the level of the Hamiltonian formulation and in 
\cite{Robinson:1994yy} at the covariant level. The almost unknown work \cite{Robinson:1994yy}
is particularly impressive for it shows how a Plebanski type Lagrangian for the 
group ${\rm U}_{\C}(2)={\rm GL}(2,\C)$ describes (complexified) unified Einstein-Maxwell theory.
The real theory can then be extracted by imposing suitable  reality conditions. 

A few years ago one of us has proposed \cite{Krasnov:2006du} a class of gravity theories 
that can be thought of as deformations of (complexified) GR in that the new theories
continue to describe, as general relativity, just two propagating DOF. These gravity 
theories are most naturally formulated using a Plebanski-type Lagrangian where the constraint
term for the two-form field is replaced by a certain potential term, see below. The 
idea of these deformations is one familiar from e.g. the relation between the scalar field $\phi^I$ theory with a Mexican hat potential and the non-linear sigma model. Here, instead of constraining $\phi^I$ to lie on the sphere $(\phi^I)^2=const$, which gives the non-linear sigma-model, one can allow the $(\phi^I)^2$ component of the field to fluctuate but give it a large mass. The original theory with the constraint $(\phi^I)^2=const$ is then recovered in the limit when
this mass is sent to infinity. Similarly, in the context of theories \cite{Krasnov:2006du} the
fields that were set to zero by the constraints of Plebanski theory are allowed to
fluctuate but these fluctuations are weighed by a potential. The original theory
corresponding to GR is recovered in the limit of an infinitely steep potential (or at low energies).
However, in the gravitational case there is one fundamental difference from the scalar field/non-linear sigma-model relationship: replacing the Plebanski constraint term by a potential does not introduce any new propagating degrees of freedom. This occurs because the fields that are allowed to fluctuate do not have kinetic terms 
and are thus non-dynamical, see e.g. \cite{Krasnov:2008fm} for a further discussion of this point. 

As it often happens, the class of theories \cite{Krasnov:2006du} has been arrived at without
knowing that it has been considered in the literature previously. Thus, 
the same class of theories (but in a rather different ``pure connection" formulation) was described and studied
a decade earlier in \cite{Bengtsson:1990qg} and following works by Bengtsson and Peldan.

The unification proposal in the context of Plebanski-type theories \cite{Krasnov:2006du} 
has been studied in \cite{Smolin:2007rx} and more recently in \cite{Lisi:2010td}. These references 
studied the unification that enlarges the gauge group of gravity formulated as a
real group ${\rm SO}(1,3)$ Plebanski-type  theory \cite{Capovilla:1991qb}.
However, in this case, once the Plebanski constraint term is replaced by a potential, 
additional propagating modes get introduced, see 
\cite{Alexandrov:2008fs} and a more recent explicit description in \cite{Speziale:2010cf}. 
Thus, if one is to avoid extra propagating modes (with some of them with wrong sign kinetic 
terms), there is no choice but to continue working in the chiral (and complex)
context of the original Plebanski theory \cite{Plebanski:1977zz}. The unification proposal
has been studied in this setting in a recent paper by two of us \cite{TorresGomez:2009gs}.

The aim of the present paper is to continue the study of unified Gravity-Yang-Mills theories
of BF type. Here we consider a particularly simple case when the underlying gauge group is
${\rm U}_\C(2)={\rm GL}(2,\C)$. We obtain a simple unified gravi-electro-magnetic
theory, and our aim is to shed some light on its properties. To this end we first look at the
pure electromagnetic sector of the model. This is obtained when  the gravitational interactions are switched off 
by setting the gravitational fields to their Minkowski spacetime values.
The resulting theory turns out to be just the most general non-linear electrodynamics with the Lagrangian being
an arbitrary function of two invariants $E^2-B^2, EB$. Such models have been studied in the
literature in the past, see in particular  \cite{Boillat:1970gw} and works by Plebanski and co-authors, including
\cite{Gutierrez:1981ed}. The usual Maxwell Lagrangian can be obtained in a limit when some parameters
of the defining function are sent to zero. 

This behaviour of the theory in the purely electromagnetic setting suggests a useful 
analogy for what happens in the case of deformations of pure GR \cite{Krasnov:2006du}.
Thus, the deformation of Einstein's GR to a more general class of theories  \cite{Krasnov:2006du} can be seen to be
analogous to the deformation of Maxwell's theory to a general class of non-linear electrodynamics 
theories. Indeed, in the Maxwell case the deformation to a non-linear electrodynamics theory consists in adding to the
Lagrangian all possible invariants constructed from the field strength. Similarly, the gravitational Lagrangian that corresponds
to deformations of GR  \cite{Krasnov:2006du} is given by a rather general expansion in invariants
constructed from the curvature, see \cite{Krasnov:2009ik}. The arising gravitational Lagrangians are
not completely general though, as the number of propagating DOF described by them is unchanged as compared to GR,
a non-trivial property that would not hold for a random Lagrangian constructed from curvature invariants.

Returning to our unified gravi-electro-magnetic theory, the general count of the degrees of freedom 
given in \cite{TorresGomez:2009gs} establishes that it is a deformation of both Einstein and Maxwell theory with the
key property that the number of propagating DOF described by this model is unchanged as compared to Einstein-Maxwell.
The deformation is controlled by a certain potential function, see below, and if one so wishes can be switched off
in a continuous fashion. Moreover, as we shall explain below, the deformation is only of significance at Planckian energies, while for low energies the theory with any generic choice of the defining potential is indistinguishable from Einstein-Maxwell.

We will pay particular attention
to the issue of how the real physical theory is extracted from an originally complex formulation,
the issue that was only touched upon in \cite{TorresGomez:2009gs}. Our present analysis of the
${\rm GL}(2,\C)$ model suggests that the prescription for dealing with reality conditions
that was advocated in \cite{TorresGomez:2009gs} must be changed. In particular, the somewhat unnatural step of 
taking the real part of the action that was used in  \cite{TorresGomez:2009gs} in an essential way is 
now eliminated altogether. This implies that the analysis of the Yang-Mills sector given in 
\cite{TorresGomez:2009gs} needs to be changed. This will be presented elsewhere.

Another possible way to think about the theories that we consider in this paper is that they arise by 
replacing the constraint term of the theory studied by Robinson \cite{Robinson:1994yy} by a potential term. Our goal is
to better understand the physics of this type of unified theories by considering the
simplest setting with a clear physical interpretation. 

Our final remarks are as follows. The natural question to confront this work with is: Why should one be interested in any deformations of Einstein-Maxwell theory? Indeed, this is a well-tested theory correctly describing our world in a vast range of scales. The first motivation is aesthetic: The two theories are quite different --- one is about spacetime metric, the other is about a ${\rm U}(1)$ connection --- and if there exists a reformulation that puts them within the same framework (at the price of allowing both theories to get deformed at very high energies) it is certainly worth studying. 

The second motivation has to do with non-renormalizability of GR. The latter implies that GR cannot be the gravity theory relevant at Planck energies. The hope is then \cite{Krasnov:2006du} that by sufficiently enlarging the class of gravitational theories one can obtain a class containing the sought ultra-violet completion of Einstein's theory. We emphasise that this differs from the usual expectation existing in the particle physics community, which is that some new DOF (possibly an infinite number of them) become relevant at Planck energies, and resolve the high energy problems of Einstein's theory. In contrast, the scenario envisaged by our approach is that it is simply the dynamics that changes as one goes to the UV, while the dynamical content of the theory (i.e. the number of propagating modes) remains unchanged. This can be seen to be a variant of the asymptotic safety scenario \cite{Weinberg:1980gg} for quantum gravity.

In the case of a unified Einstein-Maxwell theory the motivation is the same. Thus, similar to the gravitational sector scenario, the idea is that it is only the dynamics of the Maxwell part of the theory that is to change in the UV, while the dynamical content remains intact.  This is exactly what is achieved by our unified model --- our deformations of the Einstein-Maxwell theory leave the propagating mode content of the theory intact, only the dynamics is changed (in the UV region). The hope is then that this change in the dynamics may capture the non-trivial new physics occurring at high energies. We note that this motivation is similar to the original motivation behind the Born-Infeld non-linear electrodynamics \cite{Born:1934gh}.

The organization of the paper is as follows. The next section describes the class of theories
that we are going to study in this paper. In Section \ref{sec:non-lin} we switch off the gravitational
sector and study the resulting non-linear electrodynamics theory. In Section \ref{sec:spher} we switch the gravitational force back on and study the spherically-symmetric solution of the theory. We conclude with a discussion.

\section{The class of theories}

In this section we briefly desribe the class of theories that we are going to study
for an arbitrary complex Lie group $G$, and then specialize to the
case $G={\rm GL}(2,\C)$ at hand. A more detailed discussion of the general $G$
case can be found in \cite{TorresGomez:2009gs}. One first studies a complex
theory on a 4-dimensional complexified manifold, with a Lagrangian depending holomorphically on all 
the fields. At a later stage reality conditions are imposed to extract a physical real section.

\subsection{The class of theories for a general gauge group}

The main dynamical field of the theory is a Lie-algebra valued two-form $B^I$,
where the index $I$ is a Lie algebra one. The other field is a connection one-form
$A^I$. To write down the action, one needs to choose an invariant bilinear form
on the Lie algebra of $G$. When $G$ is not simple, this is not unique, and we 
assume that a choice has been made. Let us denote this form by $g_{IJ}$. The action
of the theory is then of the so-called BF type, with an extra potential term for
the $B$-field:
\be\label{action-gen}
S[B,A]=\im \int g_{IJ} B^I \wedge F^J - \frac{1}{2}V(B^I\wedge B^J).
\ee
Here $\im=\sqrt{-1}$ is a factor introduced for future convenience and
$F^I$ is the curvature two-form of the connection $A^I$. The potential term is
non-standard and needs explanation. The potential $V$ is a $G$-invariant, 
holomorphic, and homogeneous function of order one in its arguments. Thus, since its
argument is a matrix-valued 4-form $X^{IJ}=B^I\wedge B^J$, we have the following
properties of $V: V(g X g^T)=V(X), \forall g\in G; V(\alpha X)=\alpha V(X)$.
The homogeneity property is important, for it makes the potential term a scalar-valued
4-form on the manifold, which can be integrated to obtain the action. 

A somewhat more practical, but less compact form of writing the action is to introduce the
density weight one matrix:
\be
\tilde{h}^{IJ}:= \frac{1}{4}\tilde{\epsilon}^{\mu\nu\rho\sigma} B^I_{\mu\nu} B^J_{\rho\sigma}.
\ee
Then
\be\label{pot-1}
V(B^I\wedge B^J)= - V(\tilde{h}^{IJ}) \, d^4x,
\ee
where the homogeneity property makes the right-hand-side to be of the right density weight
to be integrated over the manifold. We use conventions: 
$dx^\mu \wedge dx^\nu \wedge dx^\rho \wedge dx^\sigma= 
- \tilde{\epsilon}^{\mu\nu\rho\sigma} d^4x$, which explains the minus sign in (\ref{pot-1}).
The potential function is then that of ratios of appropriate powers of all the invariants
that one can construct from $\tilde{h}^{IJ}$, see below.

The field equations that follow from (\ref{action-gen}) are as follows. Varying the
action with respect to the connection one gets:
\be
D_A B^I=0,
\ee
where $D_A$ is the covariant derivative with respect to $A$. This equation 
can often be interpreted as an equation allowing one to find the connection in
terms of derivatives of the two-form field. Varying the action with respect to $B^I$
one gets:
\be
g_{IJ} F^J = \frac{1}{2}\frac{\partial V}{\partial B^I},
\ee
which, once $A^I$ is expressed in terms of derivatives of $B^I$, 
can be interpreted as a second-order differential equation for the two-form
field. To give meaning to the right-hand-side of this equation we use the parametrisation of the potential (\ref{pot-1}). The field equation becomes:
\be
g_{IJ} F^J_{\mu\nu}  = \frac{\partial V}{\partial \tilde{h}^{IJ}} B^J_{\mu\nu}.
\ee
The derivative on the right-hand-side is now the usual derivative of a function of a matrix and can be computed without any difficulties.

\subsection{The case of $G={\rm GL}(2,\C)$}

Let us now specialize to the case of interest $G={\rm GL}(2,\C)$. The Lie algebra in
this case is 4 (complex) dimensional and splits ${\mathfrak g}={\mathfrak so}_{\C}(3)\oplus 
{\mathfrak u}_{\C}(1)$. Up to rescalings, there is a unique invariant bilinear form
in each factor. Thus, if we split $I=(i,4)$, where $i=1,2,3$ is a ${\mathfrak so}_{\C}(3)={\mathfrak su}_\C(2)$
Lie algebra index, then the most general bilinear form is:
\be
\langle X,Y\rangle = \kappa_1 \delta_{ij} X^i Y^j + \kappa_2 X^4 Y^4,
\ee
where $X^i,Y^i,X^4,Y^4$ are components of $X^I,Y^I$ and
$\delta_{ij}$ is the usual invariant form on ${\mathfrak so}_{\C}(3)$.
The curvature components are:
\be
F^i = dA^i + \frac{1}{2} \epsilon^i_{\, jk} A^j \wedge A^k, \qquad F^4 = dA^4,
\ee
where $\epsilon^i_{\, jk}$ are the ${\mathfrak so}_{\C}(3)$ structure constants. The first 
BF term of the action then takes the following form:
\be
\im \kappa_1 \int \delta_{ij} B^i\wedge F^j + \im \kappa_2 \int B^4 \wedge dA^4.
\ee
Since the normalizations of the two-form fields are not yet fixed we can freely absorb
the constants $\kappa_{1,2}$ into the fields, and we shall do so.

Let us now discuss the potential term. Let us introduce the following quantities:
\be\label{invar}
\tilde{h}^{ij} :=\frac{1}{4} \tilde{\epsilon}^{\mu\nu\rho\sigma} B^i_{\mu\nu} B^j_{\rho\sigma},
\qquad
\tilde{\phi}^{i} :=\frac{1}{4} \tilde{\epsilon}^{\mu\nu\rho\sigma} B^i_{\mu\nu} B^4_{\rho\sigma},
\qquad
\tilde{\psi} :=\frac{1}{4} \tilde{\epsilon}^{\mu\nu\rho\sigma} B^4_{\mu\nu} B^4_{\rho\sigma}.
\ee
The matrix $\tilde{h}^{IJ}$ is then:
\be
\tilde{h}^{IJ} = \left( \begin{array}{cc} \tilde{h}^{ij} & \tilde{\phi}^{j} \\
\tilde{\phi}^{i} & \tilde{\psi} \end{array} \right).
\ee
The $G$-invariants of this matrix are:
\be
{\rm Tr}(\tilde{h}), \quad {\rm Tr}(\tilde{h}^2), \quad {\rm Tr}(\tilde{h}^3),
\quad  (\tilde{\phi})^2 , \quad \tilde{\psi},
\ee
where the traces of powers of the matrix $\tilde{h}^{ij}$ are computed using the invariant
metric $\delta_{ij}$, and $(\tilde{\phi})^2=\delta_{ij} \tilde{\phi}^i \tilde{\phi}^j$.
We can take any of these quantities as the basic one, and construct
ratios of the other quantities and powers of the basic one to form quantities invariant
under rescalings of $B^I$. It is convenient to choose as the basic quantity 
${\rm Tr}(\tilde{h})$. The potential function can then be written as:
\be\label{pot}
V(\tilde{h}^{IJ}) = \frac{{\rm Tr}(\tilde{h})}{3} f\left(
\frac{{\rm Tr}(\tilde{h}^2)}{({\rm Tr}(\tilde{h}))^2}, 
\frac{{\rm Tr}(\tilde{h}^3)}{({\rm Tr}(\tilde{h}))^3},
\frac{(\tilde{\phi})^2}{({\rm Tr}(\tilde{h}))^2},
\frac{\tilde{\psi}}{{\rm Tr}(\tilde{h})}\right) \, ,
\ee
where $f$ is now an arbitrary function of its 4 arguments.

The full action, written in terms of components of forms, is:
\be\label{action-u2}
-\im \int d^4x \left( \frac{1}{4} \tilde{\epsilon}^{\mu\nu\rho\sigma}(\delta_{ij} B^i_{\mu\nu} F^j_{\rho\sigma} + 
B^4_{\mu\nu} F^4_{\rho\sigma}) - \frac{1}{2} V(\tilde{h}^{IJ}) \right) \, .
\ee
Varying this with respect to the two-form field components
one can easily obtain the field equations. It is most compact to write them using the form notations:
\be\label{eqn-bi}
F^i = \frac{\partial V}{\partial \tilde{h}^{ij}} B^j + 
\frac{\partial V}{\partial (\tilde{\phi})^2}  \tilde{\phi}^i B^4, \\ \label{eqn-b4}
dA^4 = \frac{\partial V}{\partial (\tilde{\phi})^2} 
 \tilde{\phi}^i B^i + \frac{\partial V}{\partial \tilde{\psi}} B^4,
\ee
where all partial derivatives of the potential can be obtained in an elementary
way from (\ref{pot}). We note that it might appear that a factor of two is missing from the last term on the
right-hand-side of the first equation, and the first term of the second. However, let us carefully compute the
variation. We have, dropping unessential constant factors and the integral sign:
\be\label{var-action-1}
 \frac{1}{2}  \tilde{\epsilon}^{\mu\nu\rho\sigma}(\delta B^i_{\mu\nu} F^i_{\rho\sigma} + 
\delta B^4_{\mu\nu} F^4_{\rho\sigma}) =  \frac{\partial V}{\partial \tilde{h}^{ij}} \delta \tilde{h}^{ij} +
 \frac{\partial V}{\partial (\tilde{\phi})^2 } 2 \tilde{\phi}^i \delta \tilde{\phi}^{i} + \frac{\partial V}{\partial \tilde{\psi}} \delta \tilde{\psi}.
\ee
Now, computing the variations on the right-hans-side from the definitions (\ref{invar}) we have:
\be
\delta \tilde{h}^{ij} = \frac{1}{2} \tilde{\epsilon}^{\mu\nu\rho\sigma} \delta B^{(i}_{\mu\nu} B^{j)}_{\rho\sigma}, 
\nonumber \\
\delta \tilde{\phi}^{i} =\frac{1}{4} \tilde{\epsilon}^{\mu\nu\rho\sigma} \left( \delta B^i_{\mu\nu} B^4_{\rho\sigma} +
B^i_{\mu\nu}  \delta B^4_{\rho\sigma} \right), 
\nonumber \\ \nonumber
\delta \tilde{\psi} :=\frac{1}{2} \tilde{\epsilon}^{\mu\nu\rho\sigma} \delta B^4_{\mu\nu} B^4_{\rho\sigma}.
\ee
We now substitute these into (\ref{var-action-1}) and equate to zero the coefficients in front of independent variations $\delta B_{\mu\nu}^i, \delta B^4_{\mu\nu}$. We get precisely (\ref{eqn-bi}), (\ref{eqn-b4}).

The equations obtained by varying the action with respect to
the connection components are:
\be
dB^i + \epsilon^{i}_{\,jk} A^j \wedge B^k = 0, \qquad dB^4=0.
\ee
The first equation here can be solved for the components of $A^i$ in terms of the
derivatives of $B^i$. One then substitutes the solution into (\ref{eqn-bi})
and obtains a second-order differential equation for $B^i$ involving also $B^4$.
The latter is found by integrating $dB^4=0$, and then the connection $A^4$ is found from
(\ref{eqn-b4}). Below we shall see how this procedure works explicitly by working
out the spherically-symmetric solution of our theory.

We also note that the equations of our theory are very similar to those of the unified theory 
\cite{Robinson:1994yy}, with the main difference being that the constraints $B^i\wedge B^j\sim \delta^{ij}$
and $B^i\wedge B^4=0$ of \cite{Robinson:1994yy} are absent in our case. Related to this is the absence on the right-hand-side of the Lagrange multipliers that imposed those constraints. Their role is now played by the derivatives of the potential
function. This is precisely analogous to what happens in the case of deformations of pure gravity,
where the constraint term in the action is replaced by a potential term, and the Lagrange multipliers on 
the right-hand-side of field equations for $B^i$ get replaced by $\partial V/\partial \tilde{h}^{ij}$. Thus,
the theory that we are considering is a deformation of the Einstein-Maxwell theory of precisely the
same type as the ${\rm SL}(2,\C)$-based theory with a potential is a deformation of Einstein's GR.
Similarly to the case of pure gravity, we shall see that it is possible to send some of the parameters of
the potential to infinity to recover the usual Einstein-Maxwell theory. To understand how this happens,
it is useful to first switch off the gravitational force, and consider what the theory under consideration
becomes as a purely electromagnetic theory.

\section{Non-Linear Electrodynamics}
\label{sec:non-lin}

\subsection{A version of non-linear electrodynamics}

In this section we switch off the gravitational part of the theory by fixing the ${\mathfrak so}_{\C}(3)={\mathfrak su}_{\C}(2)$
part of the 2-form field to be given by:
\be\label{sigma-sd}
B^i=\Sigma^i=\im dt\wedge dx^i-\frac{1}{2}\epsilon^i{}_{jk}dx^j\wedge dx^k\, ,
\ee
which corresponds to the Minkowski spacetime background. We further expand the $B^4$
field into the basis of self- and anti-self-dual two-forms:
\be
B^4=\Phi^i\Sigma^i+\Psi^i\bar\Sigma^i = (\Psi^i+\Phi^i) i dt\wedge dx^i +
(\Psi^i-\Phi^i) \frac{1}{2} \epsilon^{ijk} dx^j\wedge dx^k\, ,
\ee
where $\Phi^i$ and $\Psi^i$ are complex functions, and $\bar\Sigma^i$ are anti-self-dual two-forms
\be\label{sigma-asd}
\bar\Sigma^i=\im dt\wedge dx^i+\frac{1}{2}\epsilon^i{}_{jk}dx^j\wedge dx^k\, .
\ee

We now compute the action (\ref{action-u2}) on this field configuration. Using 
$\Sigma^i\wedge \Sigma^j=-2\im \delta^{ij} d^4x$ we get:
\be\label{act-phi-psi}
S[\Phi,\Psi,A^4]=\int d^4x \left( (\Phi^i\Sigma^{i\,\mu\nu}-\Psi^i\bar\Sigma^{i\,\mu\nu}) \partial_{[\mu} A^4_{\nu]} 
+f(\Phi^2,\Phi^2-\Psi^2)\right)\, ,
\ee
where $f$ is an arbitrary function of its two arguments. The action depends on fields $\Phi,\Psi,A^4$ that
are at this stage all complex. In anticipation of the reality conditions to be imposed on the connection $A^4$, let
us rewrite the Lagrangian in terms of a new connection $\bf A$:
\be
A^4=\im {\bf A} \, .
\ee
We will later require this connection to be real, with the original $A^4$ thus being an ${\mathfrak u}(1)$ connection.
Using the explicit form of (\ref{sigma-sd}), (\ref{sigma-asd}) we have:
\be
S[\Phi,\Psi,{\bf A}]=\int d^4x \left( (\Psi^i-\Phi^i) (\partial_0 {\bf A}_i -\partial_i {\bf A}_0) - \im (\Psi^i+\Phi^i)\epsilon_i{}^{jk} 
\partial_j {\bf A}_k +f(\Phi^2,\Phi^2-\Psi^2)\right).
\ee
It is now clear that the combination
$\Psi^i-\Phi^i$ plays the role of the momentum conjugate to the spatial projection of the connection $\bf A$:
\be
{\bf E}^i := \Psi^i-\Phi^i \, ,
\ee
and the combination
\be
{\bf Q}^i:=\im(\Psi^i+\Phi^i)
\ee 
is non-dynamical, to be eliminated via its field equation. The action in the Hamiltonian form thus becomes:
\be
S[{\bf E,Q,A}]=\int d^4x \left( {\bf E}^i \partial_0 {\bf A}_i + {\bf A}_0 \partial_i {\bf E}^i - {\bf Q}^i {\bf B}_i + f(({\bf E}^2-{\bf Q}^2)/4+(\im/2){\bf EQ}, \im{\bf EQ})\right),
\ee
where we have introduced the magnetic field:
\be
{\bf B}_i:=\epsilon_i{}^{jk} \partial_j {\bf A}_k.
\ee
Once the field ${\bf Q}^i$ is eliminated by solving its field equation, we get the non-linear electrodynamics action in the Hamiltonian form:
\be
S[{\bf E,A}]=\int d^4x \left( {\bf E}^i \partial_0 {\bf A}_i + {\bf A}_0 \partial_i {\bf E}^i - H( {\bf E}, \im {\bf B}) \right),
\ee
where $H$ is the Legendre transform of the original potential function $f$ with respect to the $\bf Q$
variable. Below we will see how this procedure works explicitly by working out the Lagrangian for the function
$f$ expanded in powers of its arguments. 
With the Hamiltonian $H$ being a Legendre transform of an arbitrary Lorentz-invariant function,
this is the most general non-linear electrodynamics Lagrangian, see e.g. \cite{Boillat:1970gw}, \cite{Gutierrez:1981ed}.
The only difference with the Lagrangians typically considered in the literature is
that in our case the dependence on the invariant $\bf EB$ is with a factor
of $\im=\sqrt{-1}$ in front, and so it is in general complex even after the reality condition ${\bf A, E}\in \R$ is imposed. The presence of this extra imaginary unit in the action makes the action invariant under a simultaneous operation of parity inversion and complex conjugation, similar to what happens in the case of the pure gravitational modified theory, see \cite{Krasnov:2009ik}. This is a very interesting feature of the class of theories considered, whose interpretation is still to be understood. In contrast, the non-linear electrodynamics real Hamiltonians containing odd powers of $\bf EB$ are, in general,
parity violating (if the coefficients in front of these terms are taken to be usual scalars). It is however clear that the same constraint that is imposed on the Hamiltonian of the usual non-linear
electrodynamics to have a parity-even theory in our case will produce a real Lagrangian. This will be
our strategy for dealing with reality conditions below. We leave the more interesting case of non-Hermitian Hamiltonians containing odd powers of $\bf E B$ (and its physical interpretation) to further research.

Now, to get a better insight into this theory let us consider its linearisation, in which only terms quadratic in the fields
are kept. 

\subsection{Linearised theory}

Unlike considerations of the previous subsection where we have derived the action in the Hamiltonian form
and kept ${\bf E}^i$ as an independent field, we will now integrate out all fields apart from the connection and
produce a more familiar Lagrangian that depends only on the field strength.
At the linearized level we should only keep the terms:
\be\label{f-lin}
f^{(2)}(\Phi^2,\Phi^2-\Psi^2) =\frac{\alpha}{2} \Phi^2 + \frac{\gamma}{2} (\Phi^2-\Psi^2)
\ee
in the expansion of the function $f$ in Taylor series, where $\alpha$ and $\gamma$ are constant parameters. Once this is done, we can integrate
out the fields $\Phi^i,\Psi^i$ from the action. The solutions for $\Phi^i,\Psi^i$ are given by:
\be\label{Phi-Psi-lin}
\Phi^i=-\frac{1}{\alpha+\gamma} \Sigma^{i\,\mu\nu} \partial_{[\mu} A^4_{\nu]}, \qquad
\Psi^i = -\frac{1}{\gamma} \bar\Sigma^{i\,\mu\nu} \partial_{[\mu} A^4_{\nu]} \, ,
\ee
and the resulting action is:
\be
S[A^4]=-\frac{1}{2}\int d^4x \left( \frac{1}{\alpha+\gamma} (\Sigma^{i\,\mu\nu} \partial_\mu A^4_\nu)^2 -
\frac{1}{\gamma} (\bar\Sigma^{i\,\mu\nu} \partial_\mu A^4_\nu)^2 \right).
\ee
Using the identities
\be
\Sigma^{i\mu\nu} \Sigma^{i\rho\sigma} = 2 \eta^{\mu[\rho}\eta^{\sigma]\nu} - i\epsilon^{\mu\nu\rho\sigma},
\quad 
\bar{\Sigma}^{i\mu\nu} \bar{\Sigma}^{i\rho\sigma} = 2 \eta^{\mu[\rho}\eta^{\sigma]\nu} + i\epsilon^{\mu\nu\rho\sigma}
\ee
that can be checked by an elementary computation we get:
\be
S[A^4]=\frac{1}{4} \left( \frac{1}{\gamma}-\frac{1}{\alpha+\gamma} \right) \int d^4 x\, F^{4\,\mu\nu}F^4_{\mu\nu} 
+\frac{i}{8} \left( \frac{1}{\gamma}+\frac{1}{\alpha+\gamma} \right)\int d^4 x  \, \epsilon^{\mu\nu\rho\sigma}
F^4_{\mu\nu}F^4_{\rho\sigma} \, ,
\ee
where $F^4_{\mu\nu}=\partial_\mu A^4_\nu - \partial_\nu A^4_\mu$.
Thus, modulo the (purely imaginary)  second term that is a total derivative, we get the following action:
\be
S[A^4] = \frac{\alpha}{4\gamma(\alpha+\gamma)} \int d^4 x  \, F^{4\,\mu\nu}F^4_{\mu\nu} \, .
\ee

Let us note that very little in the above analysis depends on the fact that the gravitational part of the two-form field was chosen to be (\ref{sigma-sd}). One can see that the procedure of integrating out the $B^4$ two-form field can be carried out in the same way whenever $B^i\wedge B^j\sim\delta^{ij}$. Thus, whenever the gravitational background is chosen to be "metric", in the sense that the Plebanski constraint $B^i\wedge B^j\sim\delta^{ij}$ is satisfied, it can be seen that the linearised electromagnetic Lagrangian is just the Maxwell one, with the metric being the one defined by declaring the two-forms $B^i$ to span the space of self-dual two forms. This means that the linearised electromagnetic theory is the usual Maxwell electrodynamics not only when considered around the Minkowski spacetime, but for any fixed metric background. On the other hand, when the condition $B^i\wedge B^j\sim\delta^{ij}$ is not satisfied (non-metric case using the terminology of \cite{Krasnov:2006du}), the linearised electromagnetic Lagrangian is different from that of Maxwell theory. This means that on a non-metric background light no longer has to follow geodesics of the metric defined by $B^i$. Of course, such non-metric backgrounds are only of significance in the high-energy regime (small distances). So, we can safely ignore them for low energies. Still, it would be interesting to study the effects of non-metricity on light propagation; we leave this to further research. 

\subsection{Linearised reality conditions}

Assuming (for simplicity) that both $\alpha,\gamma$ are real and positive, we easily deduce the 
linearized level reality conditions that must be imposed on our fields. Thus, the condition that
$A^4$ is purely imaginary, which is appropriate if we want to think of $A^4$ as the ${\mathfrak u}(1)$
component of a connection field, gives the correct Lorentzian signature action. Thus, for
\be\label{lin-reality}
A^4 = \im {\bf A}, \quad {\bf A}\in \R\, ,
\ee
we get:
\be
S[{\bf A}]= - \frac{1}{4 g_{{\mathfrak u}(1)}^2} \int d^4 x  \, F^{\mu\nu}F_{\mu\nu} \, ,
\ee
where $F_{\mu\nu}=\partial_\mu {\bf A}_\nu - \partial_\nu {\bf A}_\mu$ is the fields strength and the coupling constant is:
\be\label{coupl}
g_{{\mathfrak u}(1)}^2= \gamma(\alpha+\gamma)/\alpha.
\ee

We now note that, if desired, we can obtain the \cite{Robinson:1994yy} version of the electrodynamics
in which the field $B^4$ is purely anti-self-dual by sending $\alpha\to\infty$. Indeed, as clear from
(\ref{Phi-Psi-lin}), in this limit $\Phi^i$ that describes the self-dual part of $B^4$ goes to zero. In this
limit the coupling constant of our Maxwell theory becomes $g_{{\mathfrak u}(1)}^2= \gamma$. Thus, the
theory considered in \cite{Robinson:1994yy} is easily recovered.

When $\alpha\to\infty$ the two-form field $B^4$ becomes (proportional to) the anti-self-dual part of
the real field strength $F_{\mu\nu}$. In the case of finite $\alpha$ the reality conditions that $B^4$ satisfies are
much more involved. We get:
\be
\im B^4_{\mu\nu} = \frac{\alpha+2\gamma}{\gamma(\alpha+\gamma)}  F_{\mu\nu} + 
\frac{\alpha}{\gamma(\alpha+\gamma)} \frac{\im}{2} \epsilon_{\mu\nu}{}^{\rho\sigma} F_{\rho\sigma}.
\ee
The structure arising is typical for the theories under consideration in that the part of the expression that
carries the $\epsilon_{\mu\nu\rho\sigma}$ tensor contains an additional factor of $\im$ as compared to the part
that does not contain $\epsilon_{\mu\nu\rho\sigma}$.

\subsection{Non-linear electrodynamics}

Above we have analyzed the theory with the potential function $f$ truncated to its quadratic terms in
the quantities $\Phi^2,\Psi^2$. To understand the structure of the full non-linear theory we expand the
potential and keep higher powers of $\Phi^2,\Psi^2$. Thus, let us see what happens at the next order,
which is quartic (Lorentz invariance prevents us from having any cubic terms). The quartic order part of the
potential can be parametrized as:
\be\label{f-4}
f^{(4)}(\Phi^2,\Phi^2-\Psi^2) =\frac{\chi}{4} (\Phi^2)^2 + \frac{\delta}{2} \Phi^2 \Psi^2 +\frac{\xi}{4}(\Psi^2)^2 ,
\ee
where $\chi$, $\delta$ and $\xi$ are constant parameters. \\
One can now vary the action (\ref{act-phi-psi}) with respect to $\Phi^i,\Psi^i$ and solve for these fields
perturbatively in powers of $A^4$. We get for the cubic order terms:
\begin{align}
\Phi^{(3)\, i} =& \frac{1}{(\alpha+\gamma)^2} \Sigma^{i\,\mu\nu} \partial_\mu A^4_\nu
\left( \frac{\chi}{(\alpha+\gamma)^2} (\Sigma \partial A^4)^2 + \frac{\delta}{\gamma^2} (\bar\Sigma \partial A^4)^2 \right),
\\ \nonumber
\Psi^{(3)\, i} =& -\frac{1}{\gamma^2} \bar\Sigma^{i\,\mu\nu} \partial_\mu A^4_\nu
\left( \frac{\delta}{(\alpha+\gamma)^2} (\Sigma \partial A^4)^2 + \frac{\xi}{\gamma^2} (\bar\Sigma \partial A^4)^2 \right),
\end{align}
where we have introduced a compact notation
\be
(\Sigma \partial A^4)^2:=\Sigma^{i\,\mu\nu} \partial_\mu A^4_\nu \Sigma^{i\,\rho\sigma} \partial_\rho A^4_\sigma=
\frac{1}{2} F^{4\,\mu\nu}F^4_{\mu\nu} - \frac{\im}{4}\epsilon^{\mu\nu\rho\sigma} F^4_{\mu\nu} F^4_{\rho\sigma},
\ee
and similarly for $(\bar\Sigma \partial A^4)^2$. Now we can compute the quartic order Lagrangian, with the result being: 
\be
{\cal L}^{(4)}= \frac{\chi}{4(\alpha+\gamma)^4} ((\Sigma \partial A^4)^2)^2+
\frac{\delta}{2\gamma^2(\alpha+\gamma)^2}  (\Sigma \partial A^4)^2 (\bar\Sigma \partial A^4)^2+
\frac{\xi}{4\gamma^4} ((\bar\Sigma \partial A^4)^2)^2 .
\ee
This can be expanded in terms of the usual field strength invariants. Thus, using 
\be
(\epsilon^{\mu\nu\rho\sigma} F^4_{\mu\nu} F^4_{\rho\sigma})^2=-8(F^{4\,\mu\nu}F^4_{\mu\nu})^2 + 16
F^4_\mu{}^\nu F^4_\nu{}^\rho F^4_\rho{}^\sigma F^4_\sigma{}^\mu \, ,
\ee
we get the following Lagrangian:
\begin{align}\label{Lagr-4}
{\cal L}^{(4)}= &\frac{1}{16} (F^{4\,\mu\nu}F^4_{\mu\nu})^2 \left( \frac{3\chi}{(\alpha+\gamma)^4}
-\frac{2\delta}{\gamma^2(\alpha+\gamma)^2} + \frac{3\xi}{\gamma^4}\right)   \\ \nonumber
-&\frac{1}{4} F^4_\mu{}^\nu F^4_\nu{}^\rho F^4_\rho{}^\sigma F^4_\sigma{}^\mu
\left( \frac{\chi}{(\alpha+\gamma)^4}
-\frac{2\delta}{\gamma^2(\alpha+\gamma)^2} + \frac{\xi}{\gamma^4}\right)  \\ \nonumber
-&\frac{\im}{4} (F^{4\,\mu\nu}F^4_{\mu\nu})(\epsilon^{\alpha\beta\gamma\delta} F^4_{\alpha\beta} F^4_{\gamma\delta})
\left( \frac{\chi}{(\alpha+\gamma)^4}- \frac{\xi}{\gamma^4}\right) \, .
\end{align}
We can now substitute here the linearized reality conditions (\ref{lin-reality})
and obtain the Lagrangian for the real-valued connection. However, we note that now, 
unlike what happened in the quadratic order of the theory, the imaginary term in the Lagrangian
is no longer a total derivative. Thus, as we have already discussed above, 
the non-linear action for the real connection (\ref{lin-reality})
is, in general, complex. This is precisely similar to what happens in the case of the effective
gravitational Lagrangian, see \cite{Krasnov:2009ik}. There the metric Lagrangian that one gets
from a similar BF-type theory with a potential (but in the case of $G={\rm SL}(2,\C)$) at cubic order
in the curvature in general contains an imaginary term that is not a total derivative. Similar to what
we are seeing here, in the purely gravitational case it is also the higher-order interaction term
that is in general complex, while the theory linearized around the Minkowski background does not
exhibit any complexity issues. We also note that, similar to what happens in the
case \cite{Krasnov:2009ik} of pure gravity, the imaginary term is odd under parity. Thus, the full
Lagrangian is invariant under the operation of complex conjugation accompanied by parity inversion.

\subsection{Reality conditions}

There are several strategies that one could follow when facing such a non-Hermitian Lagrangian. One, advocated
in \cite{TorresGomez:2009gs}, \cite{Krasnov:2009ik} is to impose the linearized reality conditions and
take the real part of the full non-linear action. As was, however, realized more recently in the context of
work \cite{Krasnov:2010tt} on the purely gravitational theory linearized around the expanding FRW background
of relevance for cosmology, this real part of the action prescription does not in general produce a consistent theory.
In the case of the FRW background the problem arises when one considers the gravitational waves (tensor 
perturbations). 

Let us describe what the problem is in some more details. As in our electromagnetic considerations above, 
in the purely gravitational case one starts from the complex action and ``integrates
out" the non-dynamical fields to obtain an action that depends only on the physical metric.
The action one gets is a functional depending holomorphically
on the complex ``physical" fields. One has to impose some reality conditions to extract the real action. Since
the action depends on the complex fields holomorphically, one can instead consider the, say, real part
of the action, and vary it with respect to real and imaginary parts of the fields. The arising Euler-Lagrange equations
are the same as the real and imaginary parts of the complex Euler-Lagrange equations one obtains
from the holomorphic action (this follows from Cauchy-Riemann equations). Thus, the real part of the holomorphic
action considered as a functional of real and imaginary parts of all the fields carries exactly the same 
information as the original holomorphic action and can be taken as the action for the theory. However,
this action depends on twice the number of physical fields, and, moreover, kinetic terms for the
``imaginary" parts of the fields are typically negative-definite. Thus, this action describes twice the
number of propagating modes of the physical theory, and is badly unstable. Reality conditions are
needed to select a good physical sector of the theory, which describes half the modes of the complex sector
and is void of any instability problems. It is natural to require that the reality conditions one imposes are
some second-class constraints that cut the dimension of the phase space by half. However, as a consideration
of simple examples shows, for an action that is obtained as the real part of a holomorphic action, it is
in general not consistent to impose the constraint that the field is real. The reason for this is that the
condition that this constraint is preserved in time generates a secondary constraint, and the condition that
the secondary constraint is preserved in time in general produces a new constraint that is not equivalent
to the original constraint of the reality of the field. Thus, in general, requiring the field to be real imposes
more constraints than one would want. So, in general the dynamics of the Lagrangian
such as (\ref{Lagr-4}) (or the real part of this Lagrangian with all fields complexified) 
is not consistent with the reality condition that the physical field is real. If one
imposes this condition at some instant of time (and arranges the momentum to be real as well), the
dynamics will in general generate an imaginary part of the field. So, the strategy of dealing with the problem of
reality conditions for non-Hermitian Lagrangians should be more sophisticated. We will leave any attempt at such
to further research. 

For the purposes of this paper we note that, at least in our case of non-linear electrodynamics, 
we can restrict the potential function defining the theory so
that the arising Lagrangian is real (for real connections). This is precisely what is usually done in the
context of non-linear electrodynamics theories studied in the literature, where there is typically 
a restriction on the class of defining functions so that the theory is parity-invariant. Thus, 
in the case of our Lagrangian (\ref{Lagr-4}) we can arrange the coefficients in the expansion of the
function $f(\Phi^2,\Phi^2-\Psi^2)$ in such a way that the coefficient in the last term in 
(\ref{Lagr-4}) is zero. We can arrange things so that no imaginary terms
arise in the higher orders of the expansion either. This will produce a consistent theory with a real
Lagrangian as far as the electromagnetic sector is concerned. Whether a similar prescription is
possible in the gravitational sector is beyond the scope of this paper. However, in the next section we shall
see that at least in the spherically-symmetric situation the reality condition that the metric is real is completely consistent. 

\section{Spherically-symmetric solution}
\label{sec:spher}

In this section we obtain and analyze the spherically-symmetric solution of the gravi-electro-magnetic theory
described above. As we have already noted above, on non-metric backgrounds where $B^i\wedge B^j \not= \delta^{ij}$ the coupling of electromagnetism to gravity as prescribed by our theory is different that in the Maxwell case. Thus, there are two different ways that electromagnetism can be coupled to deformations of GR \cite{Krasnov:2006du}: one way is to couple the electromagnetic potential to the metric defined by declaring the gravitational sector two-forms $B^i$ to be self-dual. Such a coupling has been studied in \cite{Ishibashi:2009wj}, where also the spherically-symmetric solution was analysed. A different coupling is given by our theory. Thus, the spherically-symmetric case field equations that we shall analyse are distinct from those in \cite{Ishibashi:2009wj}. We emphasise that this difference is only of relevance for very small scales (or Planckian curvatures). For low energies (large distances) the sperically-symmetric solutions of both \cite{Ishibashi:2009wj} and this work become indistinguishable from Reissner-Nordstrom. 

\subsection{The spherically symmetric ansatz}

We start by making an ansatz for all the fields as dictated by the symmetry. The gravitational ${\mathfrak su}(2)$ sector B-fields can be selected as in the purely gravitational case
first studied in \cite{Krasnov:2007ky}. This reference has worked in spinor notations and used a complex
null tetrad. However, it is not hard to repeat the analysis for a real tetrad for the usual spherically-symmetric
metric asatz
\be
ds^2=-f^2 dt^2 + g^2 dr^2 + r^2 d\theta^2 + r^2 \sin^2(\theta) d\phi^2,
\ee
where as usual $f,g$ are (real) functions of the radial coordinate $r$ only. The starting point of the analysis
is to construct the self-dual two-forms for this metric, see e.g. \cite{Krasnov:2009pu} for a
description of this procedure for the case of Einstein's GR. The modified theory ansatz is then
obtained by allowing for an extra functions of the radial coordinate multiplying the metric B-field
ansatz of the GR case. Using the available coordinate freedom one can put the B-field in the following
convenient form:
\be\nonumber
B^1=b(ifrdt\wedge d\theta-gr\sin\theta d\phi\wedge dr), \\
B^2=b(ifr\sin\theta dt\wedge d\phi-grdr\wedge d\theta), \\ \nonumber
B^3=ifgdt\wedge dr-r^2\sin\theta d\theta\wedge d\phi \, ,
\ee
where $b$ is a function of the radial coordinate. When $b=1$ one gets the usual metric self-dual two-form
ansatz of relevance for Einstein's GR, see \cite{Krasnov:2009pu}. As we already mentioned in the previous section, when the parameters of the potential of the electromagnetic sector are chosen so that the purely electromagnetic Lagrangian is real, the metric also turns out to be real. Thus,  in the spherically-symmetric case one can assume that the metric is real from the start. We therefore assume the functions $f,g$ and $b$, as well as the coordinate functions $t,r,\theta,\phi$ to be real. 

The ansatz that we make for the $B^4$ two-form field is a general combination of the ``electric" and ``magnetic"
two-forms:
\be\label{b4}
B^4=-2 cr^2\sin\theta d\theta\wedge d\phi + 2\im f g m \, dt\wedge dr \, ,
\ee
where $c,m$ are functions of $r$ only, and the numerical constants are introduced for future
convenience. No reality conditions on $c,m$ are assumed at this stage.

\subsection{$B$-compatible ${\rm GL}(2,\C)$-connection}

We now solve $$D_AB^I=dB^I+f^I_{JK}A^J\wedge B^K=0$$ for the connection. The gravitational
${\mathfrak su}(2)$ part of this ``compatibility" equation reads:
\be
D_AB^i=dB^i+\epsilon^i_{jk}A^j\wedge B^k=0,
\ee
which gives:
\begin{align}\nonumber
A^1=&-\frac{1}{bg}\sin\theta d\phi, \\ 
A^2=&\frac{1}{bg}d\theta, \\ \nonumber
A^3=&\frac{\im f}{g}\left[\frac{(bfr)'}{bfr}-\frac{1}{b^2r}\right]dt+\cos\theta d\phi\, .
\end{align}

For the ${\mathfrak u}(1)$ part of the compatibility equation we have $D_AB^4=dB^4=0$. This implies:
\be\label{eqn-c}
(cr^2)'=0 .
\ee
Note that we cannot solve this equation for $A^4$, so we will need to find the electromagnetic connection from
another equation. For now, we make the following spherically-symmetric ansatz for it:
\be
A^4 = \im a dt + \im p \cos{\theta} d\phi \, ,
\ee
where $a$ is, at this stage, arbitrary functions of $r$, and the imaginary unit is introduced in
the expectation that later the reality condition will be imposed requiring the connection to be purely
imaginary, as appropriate for a ${\mathfrak u}(1)$ connection. The spherical symmetry requires $p$ to be a constant (proportional to the magnetic charge of our system).

We can now compute the curvature 
\be
F^I=dA^I+\frac{1}{2}f^I_{JK}A^J\wedge A^K
\ee
of the connection that we found (or made an ansatz for) above. We have for the gravitational sector $I=1,2,3$:
\be\nonumber
F^1=-\frac{1}{bg}\left\lbrace\frac{\im f}{g}\left[\frac{(bfr)'}{bfr}-\frac{1}{b^2r}\right]dt\wedge d\theta+bg\left(\frac{1}{bg}\right)'\sin\theta dr\wedge d\phi\right\rbrace , \\
F^2=-\frac{1}{bg}\left\lbrace\frac{\im f}{g}\left[\frac{(bfr)'}{bfr}-\frac{1}{b^2r}\right]\sin\theta dt\wedge d\phi-bg\left(\frac{1}{bg}\right)'dr\wedge d\theta\right\rbrace, \\ \nonumber
F^3=-\left\lbrace\frac{\im f}{g}\left[\frac{(bfr)'}{bfr}-\frac{1}{b^2r}\right]\right\rbrace'dt\wedge dr-\left(1-\frac{1}{b^2g^2}\right)\sin\theta d\theta\wedge d\phi ,
\ee
and for the electromagnetic field strength $I=4$:
\be
F^4=dA^4=-\im a' dt\wedge dr - \im p \sin(\theta) d\theta\wedge d\phi \, .
\ee

\subsection{Field Equations}

The remaining field equations to consider are given in (\ref{eqn-bi}) above. Defining the matrix $h^{IJ}$ via:
\be
B^I\wedge B^J=h^{IJ}(-2ifgr^2\sin\theta\,dt\wedge dr\wedge d\theta\wedge d\phi)
\ee
we get:
\be\label{h-mat}
h^{IJ}=\left[ \begin {array}{cccc} b^2&0&0&0\\\noalign{\medskip}0&b^2&0&0
\\\noalign{\medskip}0&0&1& c+m \\\noalign{\medskip}0&0& c+m & 4cm
\end {array} \right]
\ee
We can now compute the derivatives of the potential (\ref{pot}) needed in (\ref{eqn-bi}). We get
\begin{align}\nonumber
\frac{\partial V}{\partial h^{ij}} =& \delta_{ij} \left( \frac{f}{3} - f'_1 \frac{2(2b^4+1)}{3(2b^2+1)^2}
-f'_2 \frac{2b^6+1}{(2b^2+1)^3} - f'_3\frac{2(c+m)^2}{3(2b^2+1)^2} -f'_4 \frac{4cm}{3(2b^2+1)}\right) \\
+&f'_1 \frac{2 h_{ij}}{3(2b^2+1)} + f'_2 \frac{(h^2)_{ij}}{(2b^2+1)^2}, \\ \nonumber
\frac{\partial V}{\partial \phi^2}=&\frac{f'_3}{3(2b^2+1)}\, , \qquad \qquad \qquad \qquad \frac{\partial V}{\partial \psi}=\frac{f'_4}{3}\, .
\end{align}
Here $f'_n$ is the derivative of the function $f$ with respect to
n-th argument evaluated at $h^{ij}={\rm diag}(b^2,b^2,1), \phi^2=(c+m)^2,\psi=4cm$.

It turns out to be very convenient to separate the trace and the tracefree parts in the gravitational part,
and introduce a separate notation for the electromagnetic part potential first derivatives. Thus, let us write the
matrix of the first derivatives of the potential as:
\be
\frac{\partial V}{\partial h^{IJ}}=\left[ \begin {array}{cccc}\Lambda-\beta&0&0&0
\\\noalign{\medskip}0&\Lambda-\beta&0&0\\\noalign{\medskip}0&0&\Lambda+2\beta&{\sigma}\\\noalign{\medskip}0&0&{\sigma}&\rho
\end {array} \right]
\ee
where $\Lambda,\beta,\rho,\sigma$ are functions of $b,c,m$ given by:
\begin{align}\nonumber
\Lambda=&\frac{f}{3}- f'_1\frac{4(b^2-1)^2}{9(2b^2+1)^2} - f'_2 \frac{2(b^2-1)(b^4-1)}{3(2b^2+1)^3}
-f'_3\frac{2(c+m)^2}{3(2b^2+1)^2}-f'_4\frac{4cm}{3(2b^2+1)} , \\ \label{rho-sigma}
\beta=&f'_1\frac{2(1-b^2)}{9(2b^2+1)} + f'_2\frac{(1-b^4)}{3(2b^2+1)^2}, \\ \nonumber
\sigma=&f'_3 \frac{c+m}{3(2b^2+1)}, \qquad \qquad \qquad \qquad \rho=\frac{f'_4}{3}.
\end{align}

The field equations (\ref{eqn-bi}) then read, in the gravitational sector:
\be\label{feqs-grav-1}
-\frac{1}{bgr}\left(\frac{1}{bg}\right)'=-\frac{1}{b^2g^2r}\left[\frac{(bfr)'}{bfr}-\frac{1}{b^2r}\right]=\Lambda-\beta \, ,\\
\label{feqs-grav-2}
\frac{1}{r^2}\left(1-\frac{1}{b^2g^2}\right)=\Lambda+2\beta+2c\sigma \, , \\ \label{feqs-grav-3}
-\frac{1}{fg}\left\lbrace\frac{f}{g}\left[\frac{(bfr)'}{bfr}-\frac{1}{b^2r}\right]\right\rbrace'=\Lambda+2\beta+2m\sigma \, .
\ee
The electromagnetic sector equation gives the following two equations:
\be\label{eqs-conn}
a'=-fg(\sigma+2\rho m),\qquad i p=r^2(\sigma + 2\rho c).
\ee
Before we analyze these equations let us describe a convenient change of independent functions
that will eventually allow us to integrate the system.

\subsection{Legendre Transformation}

As was done in the purely gravitational spherically-symmetric case treated in \cite{Krasnov:2007ky},
we can think of $\Lambda$ as the Legendre transform of the function $f$. In fact, we have
\be\label{L-transf}
\Lambda=\frac{f}{3}-x\beta-y\sigma-z\rho,
\ee
where 
\be
x=2\frac{1-b^2}{2b^2+1}, \qquad y=2\frac{c+m}{2b^2+1}, \qquad z=\frac{4cm}{2b^2+1}.
\ee
Thus, we get
\be
\Lambda_\beta:=\frac{\partial\Lambda}{\partial\beta}=-x, \qquad
\Lambda_\sigma:=\frac{\partial\Lambda}{\partial\sigma}=-y, \qquad 
\Lambda_\rho:=\frac{\partial\Lambda}{\partial\rho}=-z.
\ee
We can use these relations to express the original functions $b,c,m$ appearing in our two-form
field ansatz in terms of derivatives of the new function $\Lambda=\Lambda(\beta,\sigma,\rho)$. We get:
\be\label{b}
b^2=\frac{2+\Lambda_\beta}{2(1-\Lambda_\beta)}, \qquad 
c+m=-\frac{3\Lambda_\sigma}{2(1-\Lambda_\beta)}, \qquad
2cm = -\frac{3\Lambda_\rho}{2(1-\Lambda_\beta)}.
\ee
This gives, for $c$ and $m$:
\be\label{c}
c=-\frac{3}{4}\left(\frac{\Lambda_\sigma}{1-\Lambda_\beta} +\sqrt{\frac{\Lambda_\sigma^2}{(1-\Lambda_\beta)^2}
+\frac{4\Lambda_\rho}{3(1-\Lambda_\beta)}}\right), \\ \nonumber
m=-\frac{3}{4}\left(\frac{\Lambda_\sigma}{1-\Lambda_\beta} -\sqrt{\frac{\Lambda_\sigma^2}{(1-\Lambda_\beta)^2}
+\frac{4\Lambda_\rho}{3(1-\Lambda_\beta)}}\right).
\ee
We have chosen the solution such that $m=0$ for $\Lambda_\rho=0$.
Thus, one can now take the viewpoint that the theory is parametrized by the function 
$\Lambda=\Lambda(\beta,\sigma,\rho)$, and that the above relations give us the functions
$b,c,m$ once the quantities $\beta,\sigma,\rho$ are solved for. This change of viewpoint will allow us to
integrate the field equations.

\subsection{Bianchi identities}

A very powerful method for analyzing the system of equations that we have obtained is by rewriting them
as differential equations for the functions $\beta,\sigma,\rho$. These are nothing but the Bianchi identities
obtained from the equation $DF^I=0$. Alternatively, these equations can be obtained directly from the
field equations (\ref{feqs-grav-1})-(\ref{feqs-grav-3}). Thus, differentiating the equation (\ref{feqs-grav-2}), and using one of the
equations in (\ref{feqs-grav-1}), as well as (\ref{eqn-c}) we get:
\be\label{bian-1}
\Lambda'+2\beta'+2c\sigma'=-\frac{6\beta}{r}.
\ee

Another Bianchi identity is obtained by differentiating the second equation in (\ref{feqs-grav-1}). We have:
\be\label{bian-2-t1}
\frac{1}{b^2 g^2 r} \left[ \frac{(bfr)'}{bfr} -\frac{1}{b^2 r} \right] \left( \frac{2(bg)'}{bg} +\frac{1}{r}\right) -
\frac{1}{b^2 g^2 r} \left[ \frac{(bfr)'}{bfr} -\frac{1}{b^2 r} \right] '=\Lambda'-\beta'.
\ee
We now rewrite the equation (\ref{feqs-grav-3}) expanding the terms on the left and dividing the whole equation by $b^2r$. We have:
\be\label{bian-2-t2}
\frac{1}{b^2 g^2 r} \left[ \frac{(bfr)'}{bfr} -\frac{1}{b^2 r} \right] \left( -\frac{f'}{f}+\frac{g'}{g} \right) -
\frac{1}{b^2 g^2 r} \left[ \frac{(bfr)'}{bfr} -\frac{1}{b^2 r} \right] ' =\frac{\Lambda-\beta +3\beta +2m\sigma}{b^2 r}.
\ee
Using the second equation in (\ref{feqs-grav-1}) we express $\Lambda-\beta$ in terms of other quantities and then take this term to the left-hand-side of the equation. We get:
\be\label{bian-2-t3}
\frac{1}{b^2 g^2 r} \left[ \frac{(bfr)'}{bfr} -\frac{1}{b^2 r} \right] \left( \frac{1}{b^2r}-\frac{f'}{f}+\frac{g'}{g} \right) -
\frac{1}{b^2 g^2 r} \left[ \frac{(bfr)'}{bfr} -\frac{1}{b^2 r} \right] ' =\frac{3\beta +2m\sigma}{b^2 r}.
\ee
We now subtract (\ref{bian-2-t3}) from (\ref{bian-2-t1}). We get:
\be\label{bian-2-t4}
\frac{1}{b^2 g^2 r} \left[ \frac{(bfr)'}{bfr} -\frac{1}{b^2 r} \right] \left( \frac{(b^2 fg)'}{b^2 fg} +\frac{b^2-1}{b^2r}\right)=
\Lambda'-\beta'-\frac{3\beta +2m\sigma}{b^2 r}.
\ee
We should now note that the following equation is true:
\be\label{eqn-f}
\frac{(b^2 fg)'}{b^2fg} = \frac{1-b^2}{b^2 r} \, ,
\ee
Indeed, this is just a rewrite of the first equality in (\ref{feqs-grav-1}). Therefore, the quantity in the second brackets on the left-hand-side of (\ref{bian-2-t4}) is zero and we get: 
\be\label{bian-2}
\Lambda'-\beta'=\frac{3\beta+2m\sigma}{b^2 r}.
\ee

Equations (\ref{bian-1}), (\ref{bian-2}), together with (\ref{eqn-c}), after the functions $b,c,m$ are
expressed in terms of $\beta,\sigma,\rho$ via (\ref{b}), (\ref{c}), become 3 first order differential equations for
the 3 unknown functions $\beta,\sigma,\rho$. Once these are found, the electromagnetic connection
is found from (\ref{eqs-conn}), and the metric functions $g$ is found from
\be\label{eqn-g}
\frac{1}{(bg)^2}=1-(\Lambda+2\beta+2c\sigma)r^2,
\ee
which is easily obtained from (\ref{feqs-grav-2}). The metric function $f$ is then found from (\ref{eqn-f}).

\subsection{Consistency}

Yet another Bianchi identity can be obtained from $dF^4=0$, and is equivalent to the statement that the magnetic 
charge $p=const$. On the other hand, the second equation in (\ref{eqs-conn}) expresses the magnetic charge in
terms of other functions. Differentiating this equation, and using the known from (\ref{eqn-c}) derivative of $c$ we get:
\be\label{bian-3}
\sigma'+2\rho' c + 2\sigma/r=0.
\ee
This equation can be shown to follow from the two Bianchi identities (\ref{bian-1}), (\ref{bian-2}) and the relations
(\ref{b}). Thus, let us show that (\ref{bian-3}) together with (\ref{bian-2}) imply (\ref{bian-1}). We multiply 
(\ref{bian-3}) by $2m$ and, using the last two identities in (\ref{b}) write the result as:
\be\label{bian-3'}
\frac{4m\sigma}{r}=\frac{3\rho'\Lambda_\rho+3\sigma'\Lambda_\sigma}{1-\Lambda_\beta} + 2\sigma' c.
\ee
We now use this, as well as the first identity in (\ref{b}), to write $-2b^2$ times (\ref{bian-2}) as:
\be\label{consist-1}
(\beta'-\Lambda')\frac{2+\Lambda_\beta}{1-\Lambda_\beta} + 
\frac{3\rho'\Lambda_\rho+3\sigma'\Lambda_\sigma}{1-\Lambda_\beta} + 2\sigma' c=-\frac{6\beta}{r}.
\ee
The first two terms on the left-hand-side combine to:
\be
\frac{2\beta'-2\beta'\Lambda_\beta +\Lambda'-\Lambda'\Lambda_\beta}{1-\Lambda_\beta}=\Lambda'+2\beta'.
\ee
Thus, (\ref{consist-1}) is just (\ref{bian-1}) and the obtained system of equations is consistent. 

\subsection{Non-metric gravity}

In the limit $\Lambda_\sigma=\Lambda_\rho=0$ the electromagnetic part of the theory is switched off and we recover the spherically symmetric solution \cite{Krasnov:2007ky} of non-metric gravity. The two Bianchi identities (\ref{bian-1}), (\ref{bian-2}) in this case coincide and give the following equation 
\be
(\Lambda_\beta+2)\beta'=-\frac{6\beta}{r}
\ee
for $\beta$. After this is solved the metric functions $f,g$ are determined from (\ref{eqn-g}), (\ref{eqn-f}). For more
details on the pure gravity sector solution see \cite{Krasnov:2007ky}. 

\subsection{Reissner-Nordstrom solution}

Let us now see how the usual Reissner-Nordstrom solution of GR coupled to Maxwell can be recovered. First, we
should switch off the gravity modifications, which is done by putting $\Lambda_\beta=0$ which gives $b^2=1$ and the
gravitational part of the two-form field becomes the usual spherically-symmetric triple of metric two-forms. The simplest
way to get the RN solution is to set $\Lambda_\sigma=0$ so that $m=-c$ and the $B^4$ field (\ref{b4}) is anti-self-dual.  However, let us see the appearance of the charged solution in full generality. We will also allow the magnetic charge to be
present, to illustrate how the issues of complexity should be dealt with. 

First, we need to perform the Legendre transform of the original defining function $f$. In a previous
section we have seen that in the absence of gravity modifications, and in the case which gives the usual
Maxwell theory, this function is given by (\ref{f-lin}). Inspection of (\ref{h-mat})
reveals that we have to replace $\Phi^2\to (c+m)^2, \Phi^2-\Psi^2\to 4cm$. Thus, in the case that
gives Maxwell theory our defining function is
\be
f(c,m) = \frac{\alpha}{2} (c+m)^2 + \frac{\gamma}{2} (4cm).
\ee
We then easily find $\sigma,\rho$ from (\ref{rho-sigma}):
\be\label{rho}
\sigma=\frac{\partial f}{\partial (c+m)^2} (c+m) = \frac{\alpha(c+m)}{2}, \qquad \rho=\frac{\partial f}{\partial (4cm)} = 
\frac{\gamma}{2}.
\ee
The Legendre transform (\ref{L-transf}) now gives:
\be
\Lambda = -\frac{2}{3\alpha} \sigma^2.
\ee
Note that this is independent of $\rho$, as the original function was linear in $\rho$. However, the derivative $\Lambda_\rho$
cannot be considered to be zero because it must satisfy the last equation in (\ref{b}). Thus, in this case the parametrization
by $\Lambda$ is somewhat degenerate. This can be dealt with by declaring the last equation in (\ref{b}) to be
satisfied by definition. This degeneracy is removed when one considers more complicated, non-linear dependence
on $cm$.

We can now proceed to solving the equations. We first find $\sigma$ from the second equation in (\ref{eqs-conn}). Using
the value of $\rho$ given by (\ref{rho}) we get:
\be
\sigma = \frac{\im p}{r^2} - \gamma c.
\ee
We now find $m$ from the second equation in (\ref{b}) and get:
\be
m= \frac{2 \im p}{\alpha r^2} - c \frac{2\gamma+\alpha}{\alpha} \, ,
\ee
where we have used $\Lambda_\beta=0$. 
We now use the first equation in (\ref{eqs-conn}) and, in anticipation that no modification to the electromagnetic
potential will be introduced, put $a'=-q/r^2$, where $q$ is the usual (real) electric charge. This allows us to express
the quantity $c$ in terms of $q,p$. Using $fg=1$ (which follows from (\ref{eqn-f})) we get:
\be
c= \frac{-\alpha q+(2\gamma+\alpha) \im p}{2\gamma r^2(\gamma + \alpha)} \, .
\ee
Note that this does have the required $1/r^2$ dependence on the radial coordinate. The above expression
for $c$ gives the following expression for $\sigma, m$
\be
\sigma= \frac{\alpha(q+\im p)}{2r^2(\gamma+\alpha)}, \qquad 
m = \frac{(2\gamma+\alpha) q - \alpha \im p}{2\gamma r^2(\gamma + \alpha)} \, .
\ee
We note that all quantities $\sigma, m, c$ became complex, so it is by no means obvious that one will arrive at
a real metric at the end. Note also that, interestingly, the quantities $c,m$ can be obtained one from another by
exchanging $\im p\leftrightarrow q$.

We can now solve for the unknown function $\beta$ using e.g. (\ref{bian-1}). Using 
$\Lambda'=-(4/3\alpha) \sigma\sigma'$ 
and putting all the terms not containing $\beta$ to the right-hand-side we get:
\be
\beta' = - \frac{3\beta}{r} - \frac{\alpha(q+\im p)((2\gamma+3\alpha)q-(4\gamma+3\alpha)\im p)}{6\gamma r^5 
(\alpha+\gamma)^2} \, .
\ee
The solution with correct behaviour at infinity is:
\be
\beta = \frac{r_s}{2r^3} +  \frac{\alpha(q+\im p)((2\gamma+3\alpha)q-(4\gamma+3\alpha)\im p)}{6\gamma r^4 (\alpha+\gamma)^2} \, .
\ee
Note that this quantity is, when $p\not=0$, complex even when the reality conditions are imposed.

We can finally find the metric functions from (\ref{eqn-g}). The above analysis does not seem to make it
plausible that the arising function $g$ can be real. However, once we substitute all the quantities we have
found above into (\ref{eqn-g}) we obtain:
\be
g^{-2}= 1-\frac{r_s}{r} + \frac{\alpha(q^2+p^2)}{2\gamma r^2 (\alpha+\gamma)} \, .
\ee 
Thus, the metric is the usual real Reissner-Nordstrom black hole with electric and magnetic charges provided
we choose $\alpha,\gamma$ so that:
\be
\frac{\alpha}{\gamma(\alpha+\gamma)}=2,
\ee
which is exactly the condition expected from the formula (\ref{coupl}) for the coupling constant.

To summarize, the analysis of this subsection confirms that there exist a two-parameter
family of potentials giving rise to unmodified Einstein-Maxwell system. It also illustrates how non-trivial can
the issue of reality become. Indeed, we have worked with complex quantities at
intermediate stages of the computation, but at the end all the complexity disappeared to give rise to
the real metric functions. This could have been expected from general considerations, since we have switched
off the gravitational and Maxwell sector modifications. However, it is reassuring to see this happening explicitly.

Any departure from the simple choice of the defining potential considered in
this subsection produces a modified theory, where one can either modify the gravitational sector, or
electromagnetic, or both. At this stage of the development of the theory it is unclear what constitutes a physically 
interesting potential function. Also, as we have emphasised above, our theory only becomes different from Einstein-Maxwell in the region of high curvatures. In this regime other effects (quantum mechanics) also become important. Thus, it is not at all clear if in this regime any physics can be extracted by looking at the purely classical solution. Thus, we refrain from an analysis of the modified black hole solutions at this stage. At the same time it is gratifying to know that the theory is simple enough that the problem of determining such a solution for a general defining potential reduces to three first order ODE's for the functions $\beta,\sigma,\rho$. 

\section{Discussion}

In this paper we described a simple class of unified gravi-electro-magnetic theories. The Lagrangian of our unified theory is of the BF type, with a potential for the B-field added. The gauge group on which the theory is based is (complexified) ${\rm U}(2)={\rm SU}(2)\times {\rm U}(1)$, and the ${\rm SU}(2)$ sector of our model is responsible for the gravitational interactions while the ${\rm U}(1)$ sector describes electromagnetism. 

The Lagrangian of our theory can be argued to be the most general functional of the Lie-algebra valued two-form field $B$ and the connection $A$, subject to the conditions of gauge and diffeomorphism invariance\footnote{An alternative equivalent viewpoint is obtained by integrating out the two-form field $B$ and considering the "pure connection" formulation, in which the theory becomes just the most general diffeomoprhism invariant gauge theory. We refrain from considering this point of view in the present paper.}. As such, it is a very natural starting point for a unified theory that describes both gravity and a gauge theory. It is gratifying to learn that the simplest possible nontrivial such a theory describes gravity, and the next most simple case describes a unified gravi-electro-magnetic theory. Indeed, the simplest possible gauge group that one could try to use in the general Lagrangian (\ref{action-gen}) is the Abelian ${\rm U}(1)$. However, in this case the only potential that one could write is $B\wedge B$, which produces a topological theory. Thus, the ${\rm U}(1)$ gauge group does not lead to a physically interesting theory. The next simplest group is ${\rm SU}(2)$, and for this choice we do get a non-trivial gravitational theory describing as GR just two polarizations of the graviton. Increasing the level of complexity one step further one takes the gauge group to be ${\rm U}(2)={\rm SU}(2)\times {\rm U}(1)$ studied in this paper. As we have seen, this is a unified theory of electromagnetism and gravity. It is quite encouraging that the first two simplest non-trivial cases that one encounters when studying diffeomorphism invariant gauge theories (\ref{action-gen}) are the two most physically important theories. 

We would like to emphasise that the only assumptions that were used in our choice of the Lagrangian are those of diffeomorphism and gauge invariance.  It is true that the potential function in (\ref{action-gen}) can be arbitrary, so one has not a single theory, but the whole class parameterised by the choice of $V(\cdot)$. However, we have also seen that for {\it any choice} of the potential the theory expanded around the Minkowski spacetime background is for low energies indistinguishable from the Einstein-Maxwell theory. Indeed, in the lowest, quadratic order in the expansion the gravitational and electromagnetic parts are uncoupled. One can then refer to the pure gravity case results in e.g. \cite{TorresGomez:2009gs} to see that the low-energy gravitons are unmodified.  On the other hand, the electromagnetic sector is described at low energies by the quadratic terms (\ref{f-lin}) which we have seen to lead to the usual Maxwell Lagrangian. Thus, for any choice of the potential function defining the theory (apart from that giving the topological BF theory) the linearized excitations are the usual gravitons and photons. One can also see that the interactions between these are correctly described by noting, as we already did above, that on an arbitrary metric background described by two-forms satisfying $B^i\wedge B^j\sim \delta^{ij}$ the linearised electromagnetic sector coupling to gravity is just that of the Maxwell theory. Thus, for low electromagnetic field strength and small curvatures in the gravitational sector our theory is just the Maxwell theory coupled to gravity. 

Thus, most of the parameters of the potential function $V$ are invisible at low energies, and we only have access to the usual Newton's constant and the Maxwell theory coupling constant (unmeasurable unless a coupling to matter is introduced). The departure of our theory from Einstein-Maxwell would only be visible at high energies (or high field strength). Since the only dimensionful parameter in the theory is the Newton's constant, the only relevant energy scale is the Planck scale. Thus, the departure of our unified theory from Einstein-Maxwell is only going to be visible close to the Planckian domain. For low energies  that we have access to any choice of $V$ leads to the usual Einstein-Maxwell system. 

We could also have phrased the above discussion in terms of renormalisation group flow arguments. Since our theory was seen to describe the degrees of freedom of the Einstein-Maxwell theory, and has the same symmetries, we are guaranteed that at low energies the theory will be indistinguishable from Einstein-Maxwell, since by general arguments the Einstein and Maxwell Lagrangians are the only terms of the effective field theory Lagrangian of gravity and electromagnetism that survive at low energies. This is confirmed by our analysis.

The choice of the potential therefore only matters in the high energy regime where the theory starts to deviate from its low energy Einstein-Maxwell limit. However, here other effects are becoming important as well, of which the most important is of course quantum mechanics. Indeed, at high energies we can no longer treat our theory as classical. This in particular implies that we are no longer free to specify the potential $V$, as it will become running with the energy scale. So, at high energy where a choice of $V$ would be of importance we are no longer free to make this choice, and it is the renormalisation group flow which will determine $V$. This is gratifying, because it means that there is no freedom at all in our choice of the Lagrangian, and it is completely specified by symmetries and the quantum behaviour. Unfortunately, we are not able to say anything about the latter at this stage of the development of the theory. 

Let us finish with a quick discussion of the open problems of this approach. First, while there is some scope for gauge-gravity unification in the context of our diffeomorphism invariant gauge theories, there are other field species in Nature - fermions. These usually require a metric (or a tetrad) if they are to couple to gravity, and so it is not at all clear that they will be possible to describe in an approach that trades the spacetime metric for a collection of two-forms. This is a difficult question that will be dealt with elsewhere. One other very important question is that of reality conditions for the whole theory (including the gravitational sector). There are some puzzles here associated with the fact that the Lagrangians one naturally obtains in our approach are non-Hermitian (this non-Hermiticity is only visible in higher-order interaction terms, see e.g. the last term in (\ref{Lagr-4}), and is of no significance at low energies). We will return to this problem elsewhere. For now let us just mention the fact that it is now known that the condition of Hermiticity is overly restrictive and there are some systems in Nature that are described by non-Hermitian Hamiltonians, see \cite{Bender:2007nj} for a review. Finally, the most important open problem of the whole approach is to study the quantum mechanical behaviour of our theories and show that they continue to make sense also as quantum theories. \\

\noindent{\bf Acknowledgements.} ATG and CS were supported by a Mathematical Sciences Research Scholarship.

\end{document}